\documentclass[seceq]{ptptex}

\usepackage{graphicx}
\usepackage{wrapft}

\title{
Chiral Symmetry in Strongly Interacting Matter%
}
\subtitle{From Nuclear Matter to Phases of QCD }    

\author{
Wolfram \textsc{Weise}%
}

\inst{
Physik-Department, Technische Universit\"at M\"unchen\\D-85747 Garching, Germany
}

\abst{
This is a brief summary of topics that were presented as lectures within the programme ``New Frontiers in QCD 2010'' at the Yukawa Institute of Theoretical Physics in Kyoto. The basic subject is phases and symmetry breaking patterns as they emerge from the approximate chiral symmetry of QCD.  Part I focuses on the QCD interface with nuclear physics via chiral effective field theory. This includes nuclear thermodynamics and, in particular, constraints for compressed and hot baryonic matter
provided by the density and temperature dependence of the chiral condensate. Part II explores aspects of the QCD phase diagram  using a non-local improved version of the Polyakov -- Nambu -- Jona-Lasinio (PNJL) model. A prominent feature of such an approach is the occurence of a dynamical entanglement between chiral and deconfinement crossover transitions. Comparisons with available results from lattice QCD thermodynamics will be made. 
}

\begin{document}

\maketitle

\section{Part I: Nuclear Chiral Dynamics}

Chiral effective field theory is the low-energy realization of QCD in the meson and single-baryon sectors. It is also an appropriate framework for dealing with the nuclear many-body problem in terms of in-medium chiral perturbation theory \cite{FKW2005}. In this approach, chiral one- and two-pion exchange processes in the nuclear medium are treated explicitly while unresolved
short-distance dynamics are encoded in contact interactions. Three-body forces emerge naturally and play a significant role in this framework. The pion mass $m_\pi$, the nuclear Fermi momentum $p_F$ and the mass splitting $M_\Delta - M_N \simeq 2 m_\pi$  between the $\Delta(1232)$ and the nucleon are all comparable ``small" scales that figure as expansion parameters. The relevant,
active degrees of freedom at low energy are therefore pions, nucleons and $\Delta$ isobars. Two-pion exchange interactions produce intermediate-range Van der Waals - like forces involving the large spin-isospin polarizablity of the individual nucleons. The Pauli principle plays an important role acting on intermediate nucleons as they propagate in two-pion exchange processes within the nuclear medium. 

\subsection{In-medium chiral perturbation theory}

A key ingredient of this approach is the in-medium nucleon propagator,
\begin{equation}
{i(\gamma_\mu p^\mu - M_N + i\varepsilon})^{-1} - 2\pi(\gamma_\mu p^\mu + M_N)\,\delta(p^2 - M_N^2)
\,\theta(p^0)\,\theta(p_F -|\vec{p}\,|)~,
\end{equation}
that takes into account effects of the filled nuclear Fermi sea (with Fermi momentum $p_F$). These propagators enter in the loop diagrams generating the free energy density. Thermodynamics is introduced using the Matsubara formalism. The free energy density is then computed as a function of temperature $T$ at given Fermi momentum $p_F$. The loop expansion of this in-medium chiral perturbation theory is connected in a one-to-one correspondence to a systematic expansion of the energy density in powers of $p_F$, with expansion coefficients expressed as functions of the dimensionless ratio $m_\pi/p_F$.  

Calculations have been performed up to and including three-loop order in the energy density.
A very limited set of constants associated with $NN$ contact terms is fixed by reproducing e.g. the binding energy per nucleon in equilibrium nuclear matter at $T=0$. Then this scenario leads to a realistic nuclear matter equation of state \cite{FKW2005} with a liquid-gas first order phase transition and a critical temperature of about 15 MeV (see Fig.1), close to the range of empirical values extracted for this quantity. Incidentally, this nuclear liquid-gas transition is so far the only well established part of the phase diagram for strongly interacting matter at finite density, with the possible exception of the chemical freezeout curve\cite{ABS2009} deduced from hadron production in heavy-ion collisions. 

The three-loop truncation in the chiral expansion of the pressure in powers of the Fermi momentum, $p_F$, implies that these calculations can be safely trusted up to about twice the density of normal nuclear matter, i.e. for $p_F \lesssim 0.3$ GeV $<< 4\pi f_\pi$ with $f_\pi = 0.09$ GeV the pion decay constant in vacuum. 

\begin{figure}[htb]
\begin{minipage}[t]{7cm}
\includegraphics[width=6.2cm]{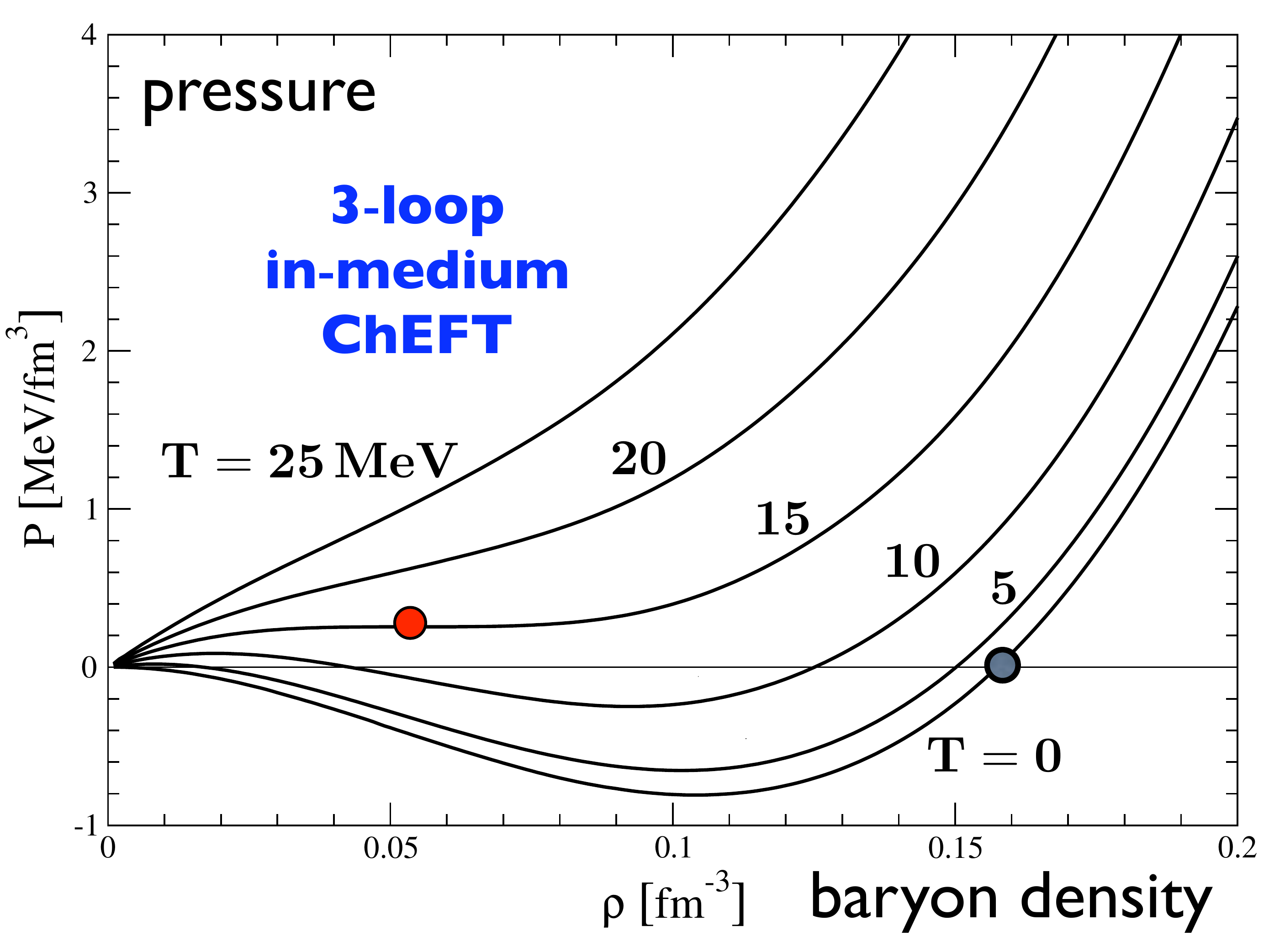}
\label{fig:1}
\end{minipage}
\hspace{\fill}
\begin{minipage}[t]{7cm}
\includegraphics[width=6.5cm]{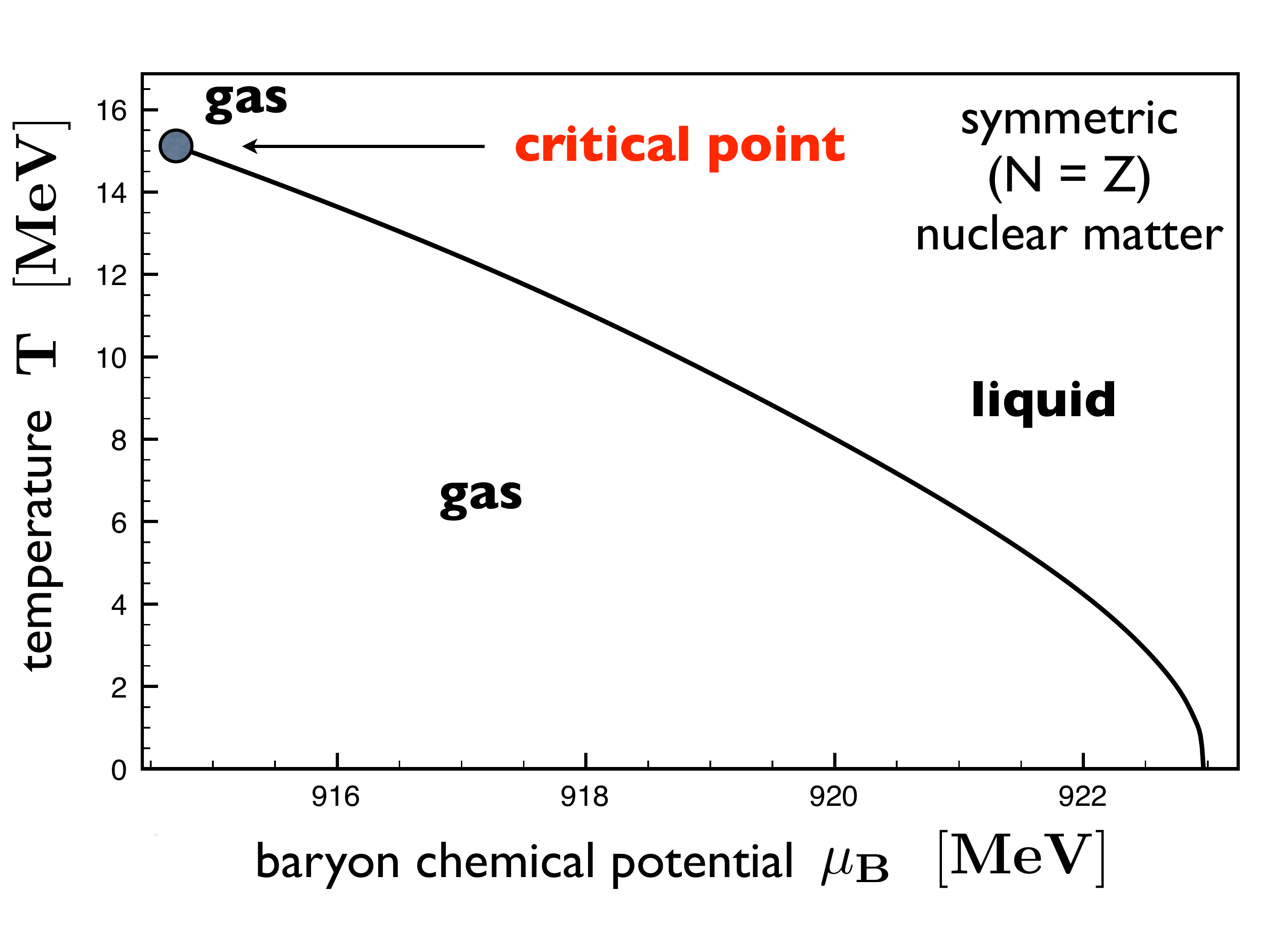}
\end{minipage}
\caption{Equation of state (left) and phase diagram (right) of symmetric $(N=Z)$ nuclear matter from in-medium chiral effective field theory \cite{FKW2005, FKW2010}. The critical point for the first-order liquid-gas phase transition appears at $T_c \simeq 15$ MeV. The equlibrium point at $T = 0$ and baryon density $\rho_0 \simeq 0.16$ fm$^{-3}$ is fixed by a single parameter, the strength of an NN contact interaction. All remaining input is pre-determined by known pion-nucleon interactions in vacuum.} 
\end{figure}
%

Figure 2 shows ($T,\rho$) phase diagrams for isospin-asymmetric nuclear matter, outlining the evolution of the critical point and the gas-liquid coexistence region for increasing proton fractions $Z/A$. The first order phase transition stops existing at $Z/A=0.05$, at which point the liquid component disappears and neutron matter begins to be realized as an interacting Fermi gas. It is important to note that this behaviour is almost entirely driven by the isospin dependence of the in-medium two-pion exchange interactions. Typical ingredients of one-boson exchange phenomenology (such as rho and sigma bosons) do not appear in this approach and are replaced by explicit $2\pi$ exchange mechanisms.

\begin{figure}[htb]
\centerline{\includegraphics[width=7.5cm,height=5.5cm] {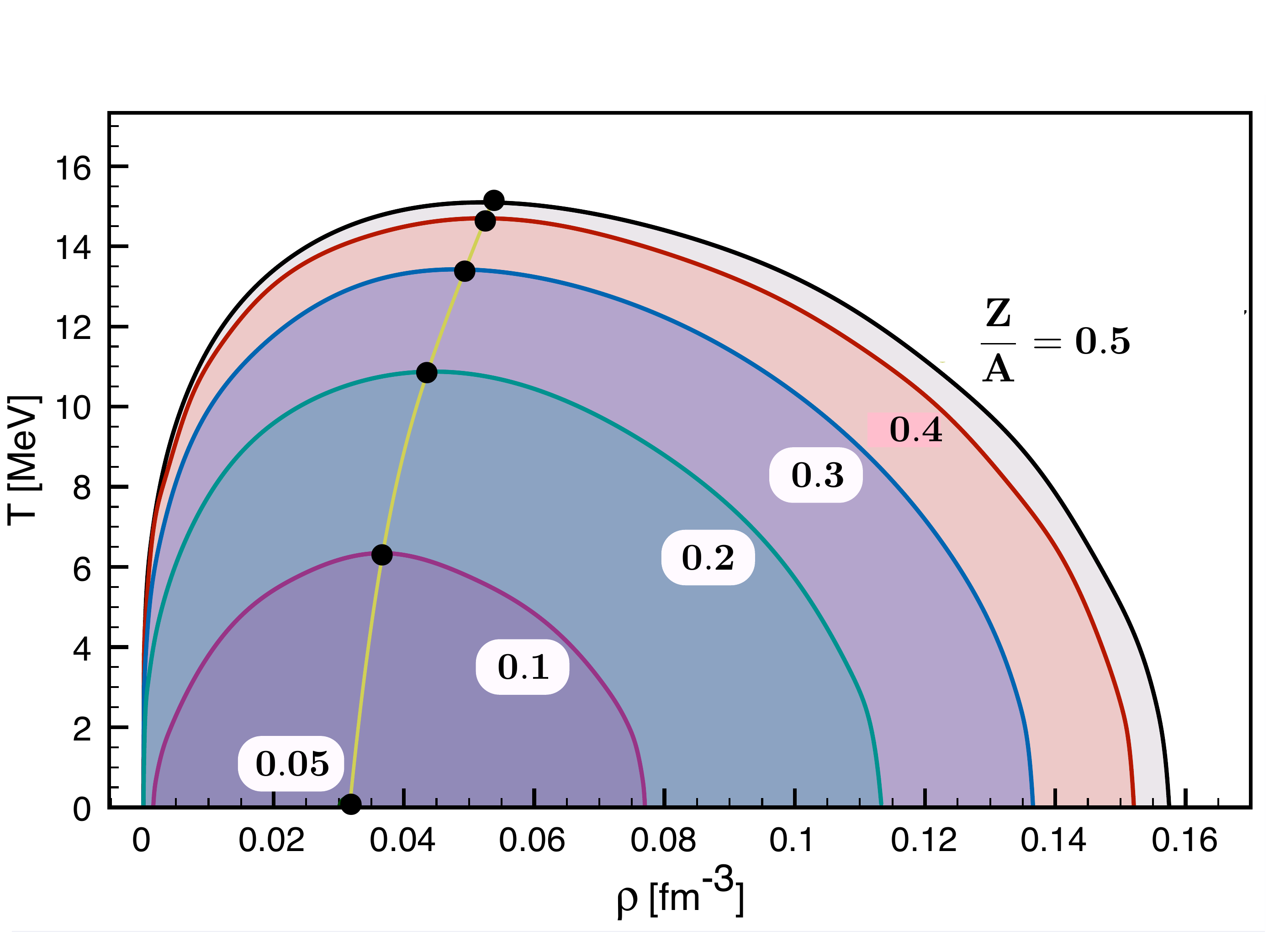}}
\label{fig:2}
\caption{Phase diagram for isospin-asymmetric nuclear matter from in-medium chiral effective field theory \cite{FKW2010} plotted in the $(T,\rho)$ plane with $\rho = \rho_p + \rho_n$ as function of the proton fraction $Z/A$. The trajectory of the critical point is also shown.}
\end{figure}
%

Extensions of this framework to finite nuclei, using the energy density functional formalism, have been
applied successfully \cite{FKVW2006} to calculate ground state properties (binding energies, mean square radii, deformations and systematics along isotopic chains) throughout the nuclear chart,  from $^{16}O$ to $Pb$ isotopes. Other examples of detailed nuclear structure phenomena such as the anomalously weak $^{14}C$ beta decay transition, with its lifetime of almost six thousand years that enables radiocarbon dating, can be understood in terms of the density dependent chiral effective interaction including a pronounced three-body contribution \cite{HKW2009}.

\subsection{Density and temperature dependence of the chiral condensate}

In a nuclear equation of state based on chiral dynamics the pion mass enters explicitly (or, equivalently, the quark mass $m_q$ according to the Gell-Mann - Oakes - Renner relation,  $m_\pi^2 f_\pi^2 = -m_q\langle\bar{\psi}\psi\rangle$). Equiped with such an equation of state one can now ask the following questions: how does the chiral (quark) condensate $\langle\bar{\psi}\psi\rangle$ extrapolate
to baryon densities $\rho$ exceeding those of normal nuclear matter, and how does it evolve as a function of temperature $T$ in a nuclear environment? The answer, first at $T=0$, is based on the Hellmann-Feynman theorem applied to the Hamiltonian density of QCD in the nuclear ground state $|\Psi\rangle$, 
\begin{equation}
\langle\Psi|\bar{\psi}\psi|\Psi\rangle = \langle\Psi|{\partial{\cal H}_{QCD}\over\partial  m_q}|\Psi\rangle = \langle\bar{\psi}\psi\rangle_0\left[1 - {1\over  f_\pi^2} {\partial{\cal E}(m_\pi;\rho)\over\partial m_\pi^2}\right]~~.
\label{eq:condensate}
\end{equation}
where $\langle\bar{\psi}\psi\rangle_0$ is the chiral condensate in vacuum.  The energy density ${\cal E}$ (normalized so that it vanishes at $\rho \rightarrow 0$) is expressed at given baryon density $\rho$ as a function of the
pion mass. In-medium chiral effective field theory gives the following result:
\begin{eqnarray}
{\langle\bar{\psi}\psi\rangle_\rho\over\langle\bar{\psi}\psi\rangle_0} = 1&-&{\rho\over f_\pi^2}{\sigma_N\over m_\pi^2}\left(1 - {3\,p_F^2\over 10\,M_N^2} + \dots\right) - {\rho\over f_\pi^2}{\partial\over \partial m_\pi^2}\left({E_{int}(m_\pi; p_F)\over A}\right)~. 
\end{eqnarray}
The second term on the r.h.s., with its leading linear dependence on density, is the contribution from a free Fermi gas of nucleons. It involves the pion-nucleon sigma term $\sigma_N\simeq 0.05$ GeV and additional non-static corrections. The third term includes the pion mass dependence of the interaction energy per nucleon, $E_{int}/A$. This term features prominently the two-pion exchange interaction in the nuclear medium  including Pauli principle corrections, and also three-nucleon forces based on two-pion exchange.

\begin{figure}[htb]
\begin{minipage}[t]{7cm}
\includegraphics[width=6.2cm]{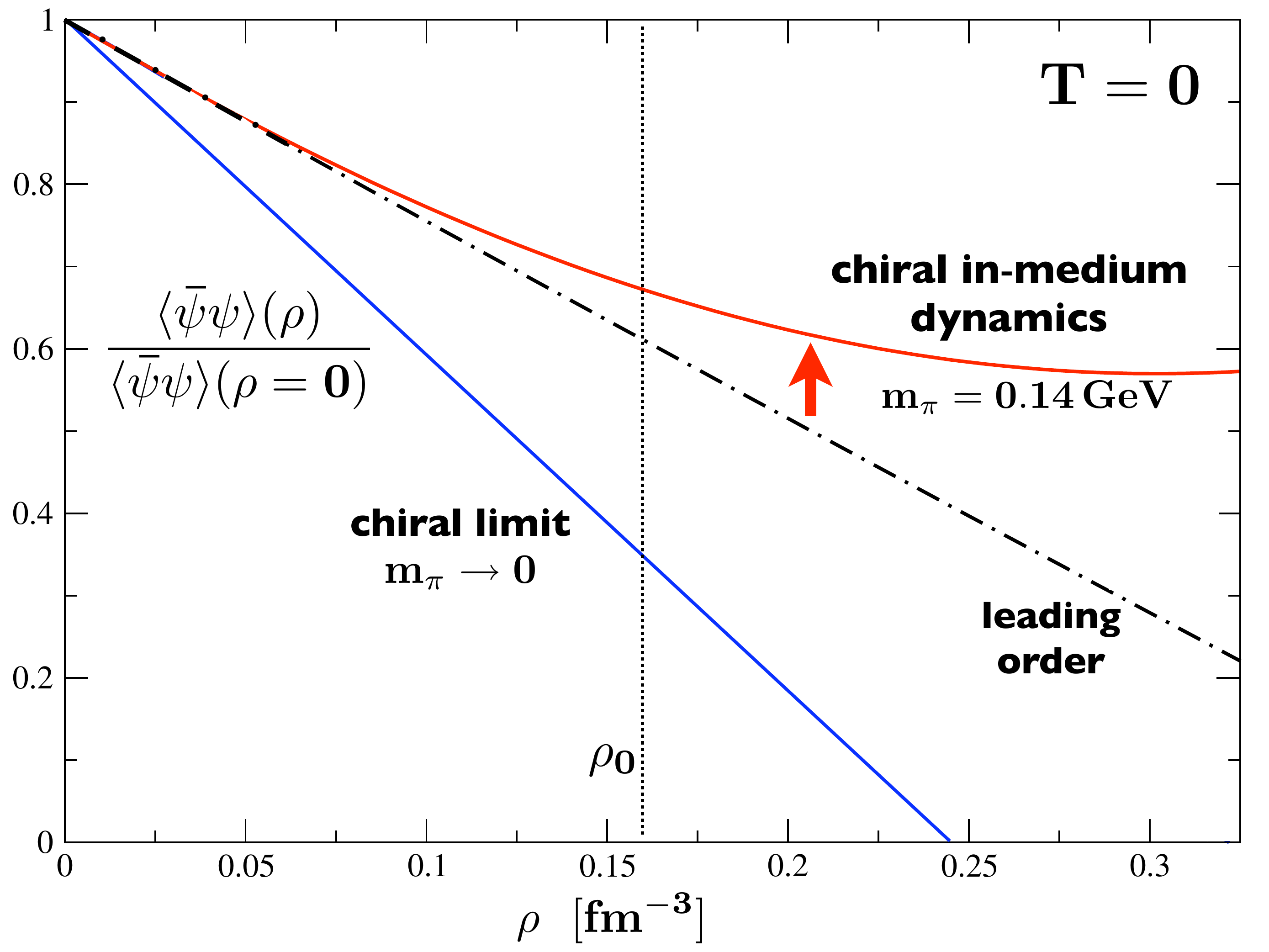}
\end{minipage}
\hspace{\fill}
\begin{minipage}[t]{7cm}
\includegraphics[width=6.5cm]{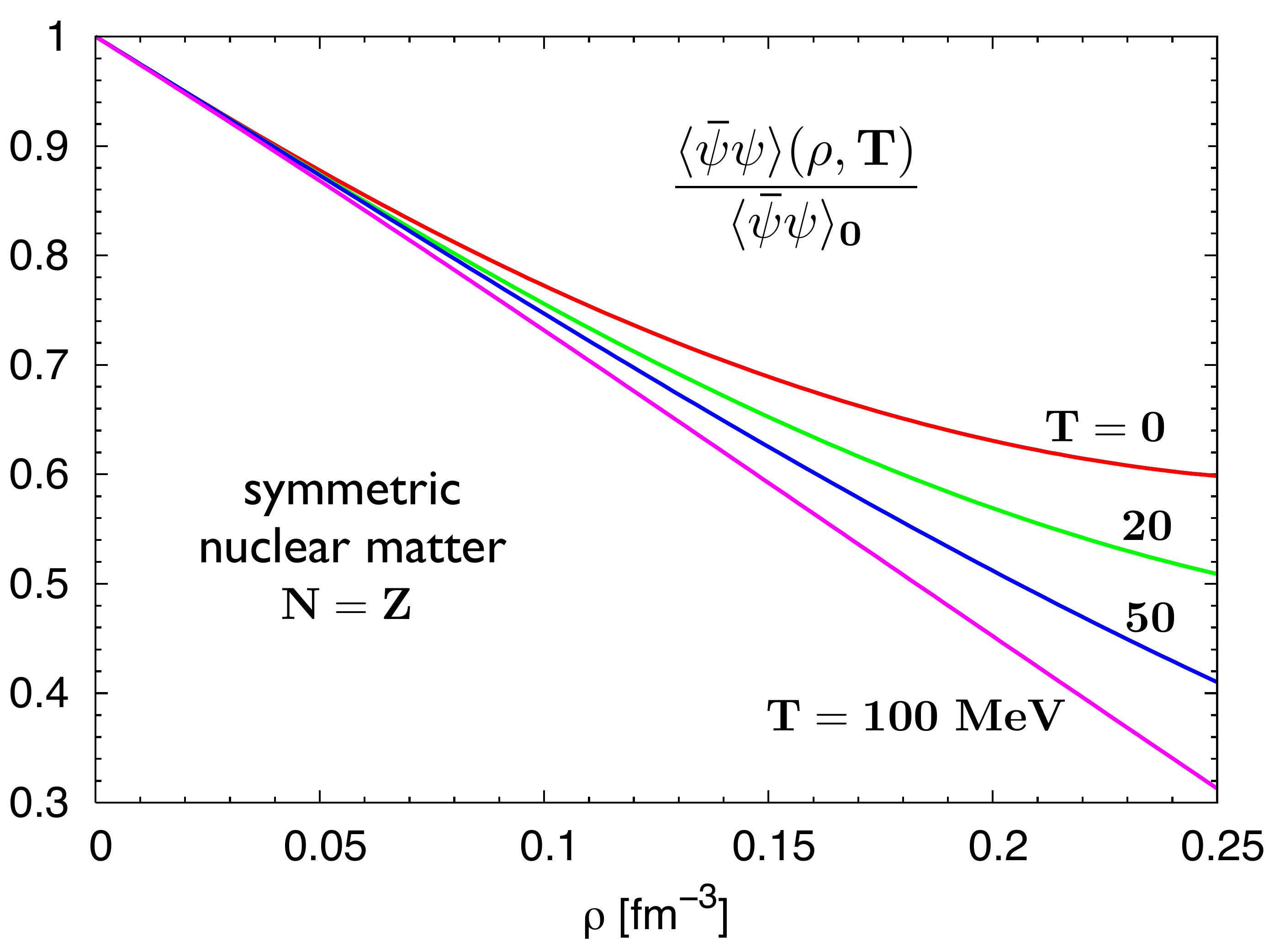}
\end{minipage}
\caption{Left: density dependence of the chiral condensate at temperature $T = 0$ in symmetric nuclear matter \cite{KHW2008}. Dot-dashed curve: leading order term using $\sigma_N = 50$ MeV. Upper curve: full in-medium chiral effective field theory  result at three-loop order. Lower curve: chiral limit with vanishing pion mass. Right: temperature dependence of the chiral condensate starting from the upper curve of the left figure at $T=0$ and using finite-temperature in-medium chiral perturbation theory \cite{FKW2010}.} 

\end{figure}
%

The dashed-dotted curve in Fig.3 (left) shows the pronounced leading linear reduction in the magnitude of the chiral condensate with increasing density. This holds in the absence of correlations between the nucleons. Up to about the density of normal nuclear matter, this term dominates, whereas the interaction part tends to delay the tendency towards chiral restoration when the baryon density is further increased. 
This stabilizing effect arises from the combination of Pauli blocking in two-pion exchange processes and
three-body correlations, and their explicit dependence on the pion mass. This behaviour is highly sensitive to the actual value of the pion mass. In the chiral limit, $m_\pi\rightarrow 0$, with stronger attraction in the NN force at intermediate ranges, the trend is reversed and the more rapidly dropping condensate would now lead to the restoration of chiral symmetry at relatively low density. In essence, nuclear physics would look radically different if pions were exactly massless Nambu-Goldstone bosons.  The drastic influence of explicit chiral symmetry breaking in QCD through the small but non-zero $u$ and $d$ quark masses on qualitative properties of nuclear matter is quite remarkable. 

The additional temperature dependence of the in-medium chiral condensate is found by repeating the calculation, Eq.(\ref{eq:condensate}), using the free energy density, ${\cal F}(m_\pi; \rho, T)$:
\begin{equation}
{\langle\bar{\psi}\psi\rangle_{\rho, T}\over \langle\bar{\psi}\psi\rangle_0} = 1 - {1\over  f_\pi^2} {\partial{\cal F}(m_\pi;\rho, T))\over\partial m_\pi^2}~~.
\end{equation} 
The result shown in Fig. 3 (right) demonstrates how the chiral condensate ``melts" with increasing temperature and approaches the linear density dependence of the non-interacting Fermi gas at a temperature of about 100 MeV. The effect of thermal pions on the temperature dependence of the quark condensate must still be added; it amounts \cite{GL89, Kai99} to lowering the magnitude of the condensate by another 5\% of its vacuum value at $T\simeq 100$ MeV. However, no tendency towards a first order chiral phase transition
is visible, at least up to baryon densities about twice that of normal nuclear matter and up to temperatures $T \sim 100$ MeV. In that range it appears that the relevant fermionic quasiparticle degrees of freedom are indeed nucleons rather than quarks. This should be kept in mind as an important constraint for the discussion of the QCD phase diagram at finite baryon chemical potential.  

\section{Part II: Phases of QCD - chiral and deconfinement transitions}

Confinement and spontaneous chiral symmetry breaking in QCD are governed by  two basic symmetry principles:

{\cal i)} The symmetry associated with the center $Z(3)$ of the local $SU(3)_c$ color gauge group is exact in the limit of pure gauge QCD, realized for {\it infinitely heavy} quarks. In the high-temperature, deconfinement phase of QCD this $Z(3)$ symmetry is spontaneously broken, with the Polyakov loop acting as the order parameter.

{\cal ii)} Chiral $SU(N_f)_R\times SU(N_f)_L$ symmetry is an exact global symmetry of QCD with $N_f$ {\it massless} quark flavors.  In the low-temperature (hadronic) phase this symmetry is spontaneously broken down to the flavor group $SU(N_f)_V$ (the isospin group for $N_f = 2$ and the ``eightfold way" for $N_f = 3$). As a consequence there exist $N_f^2 - 1$ pseudoscalar Nambu-Goldstone bosons and the QCD vacuum hosts a strong quark condensate. 

There is no principal reason why the chiral and deconfinement transitions should be closely connected. Nonetheless this appears to be the case in lattice QCD computations with almost physical quark masses.  Examples of results from lattice QCD thermodynamics \cite{Ch2008} with 2+1 flavors (at zero baryon chemical potential) are shown in Fig. 4. More recently, improved computations\cite{Bo2010, BP2010} (see Fig. 5) indicate a shift of the chiral transition to somewhat lower temperatures, consistent with earlier lattice simulations \cite{Aoki2006}. Nevertheless, given the smoothness of the crossover, the chiral and deconfinement transitions appear to be linked and overlapping at zero chemical potential, $\mu = 0$.  
\begin{figure}[htb]
\begin{minipage}[t]{7cm}
\includegraphics[width=6.5cm]{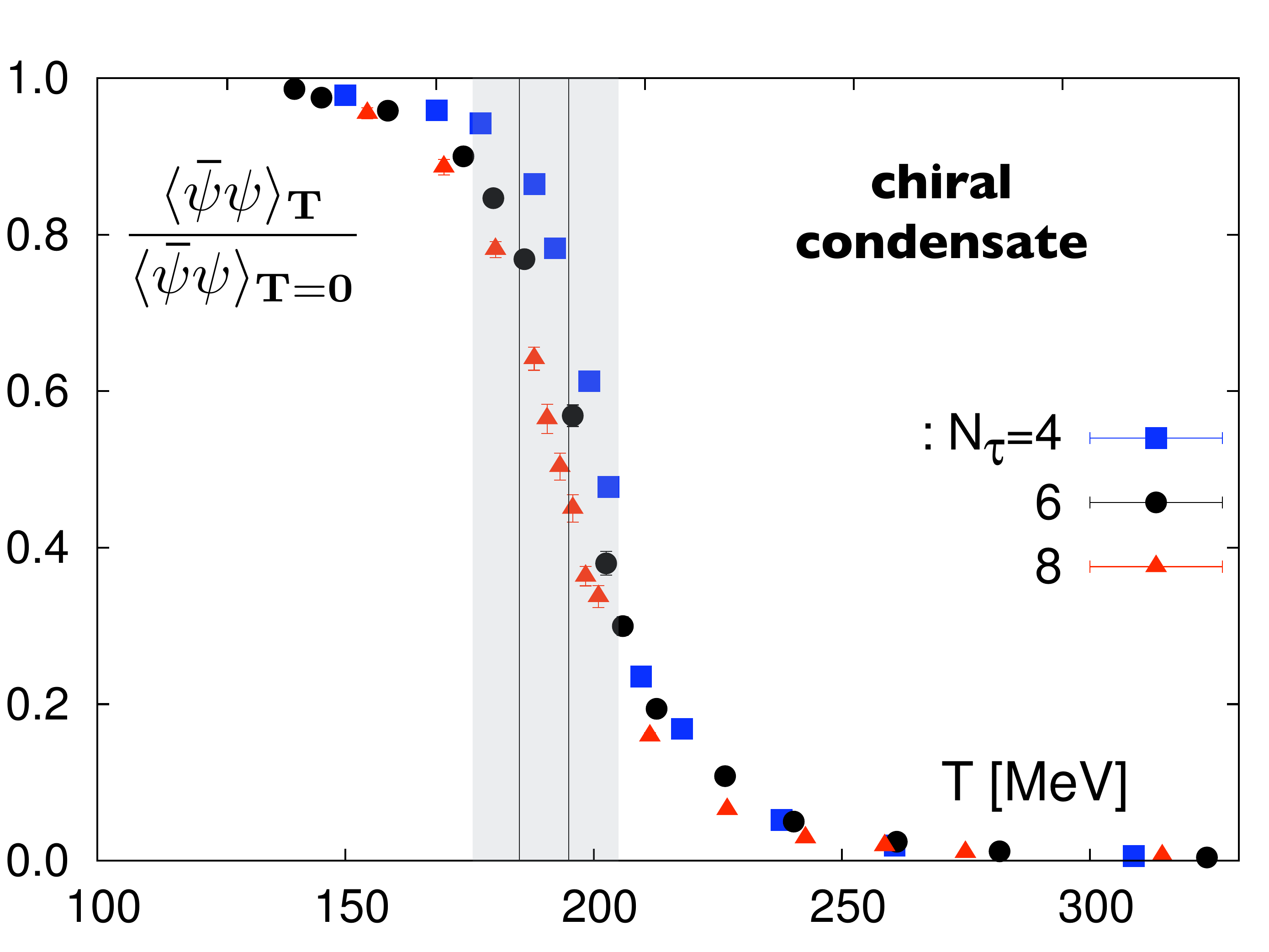}
\end{minipage}
\hspace{\fill}
\begin{minipage}[t]{7cm}
\includegraphics[width=6.5cm]{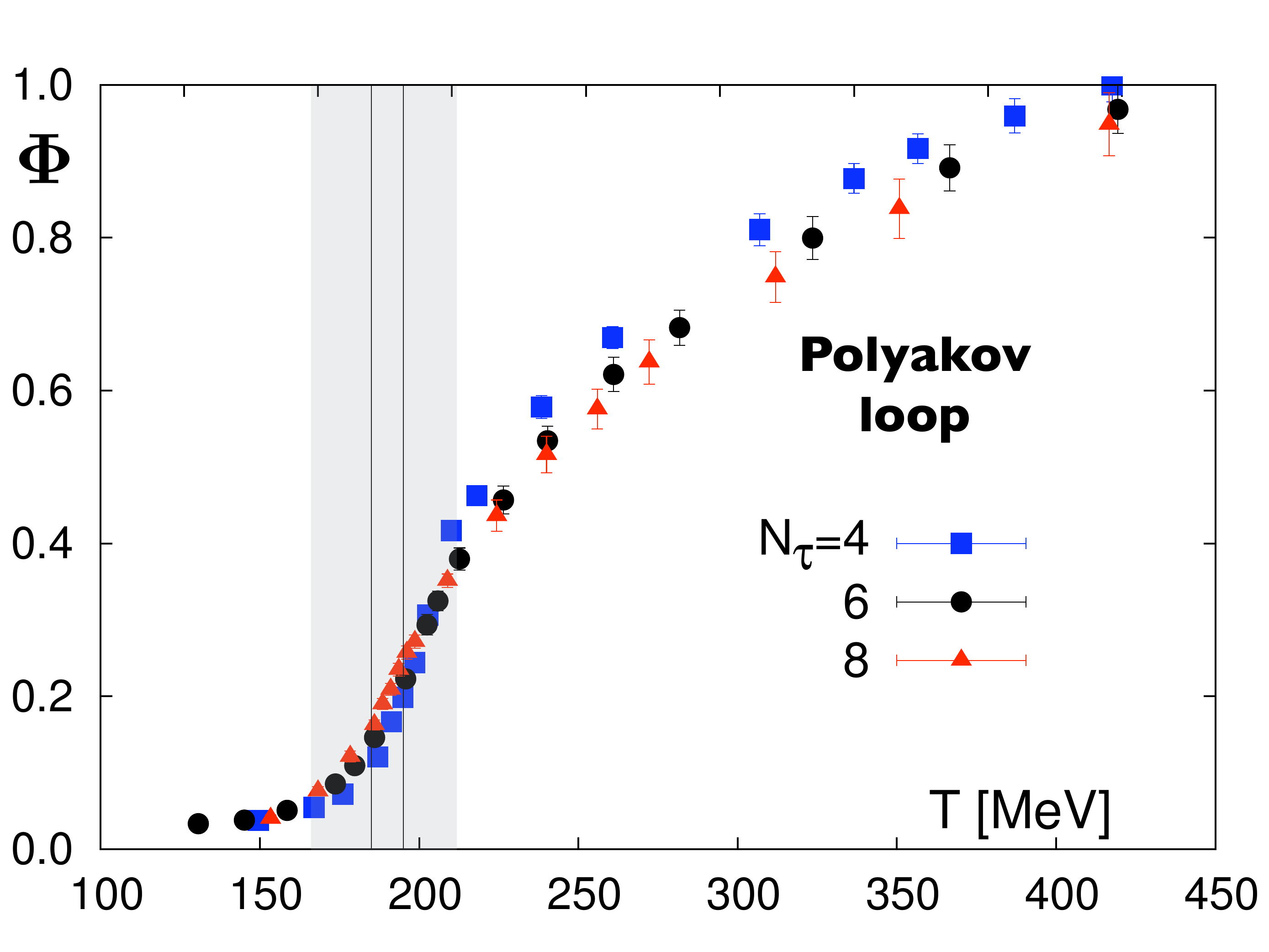}
\end{minipage}
\caption{Lattice QCD results ($N_f = 2+1$) of chiral condensate (left) and Polyakov loop (right) as functions of temperature \cite{Ch2008}. The grey bands mark the crossover regions and give an impresion of the transition temperature range. Different data sets correspond to different number $N_\tau$ of lattice points along the Euclidean time axis.} 
\label{fig:4}
\end{figure}
%

\begin{figure}[htb]
\centerline{\includegraphics[width=6.5cm,height=5.5cm] {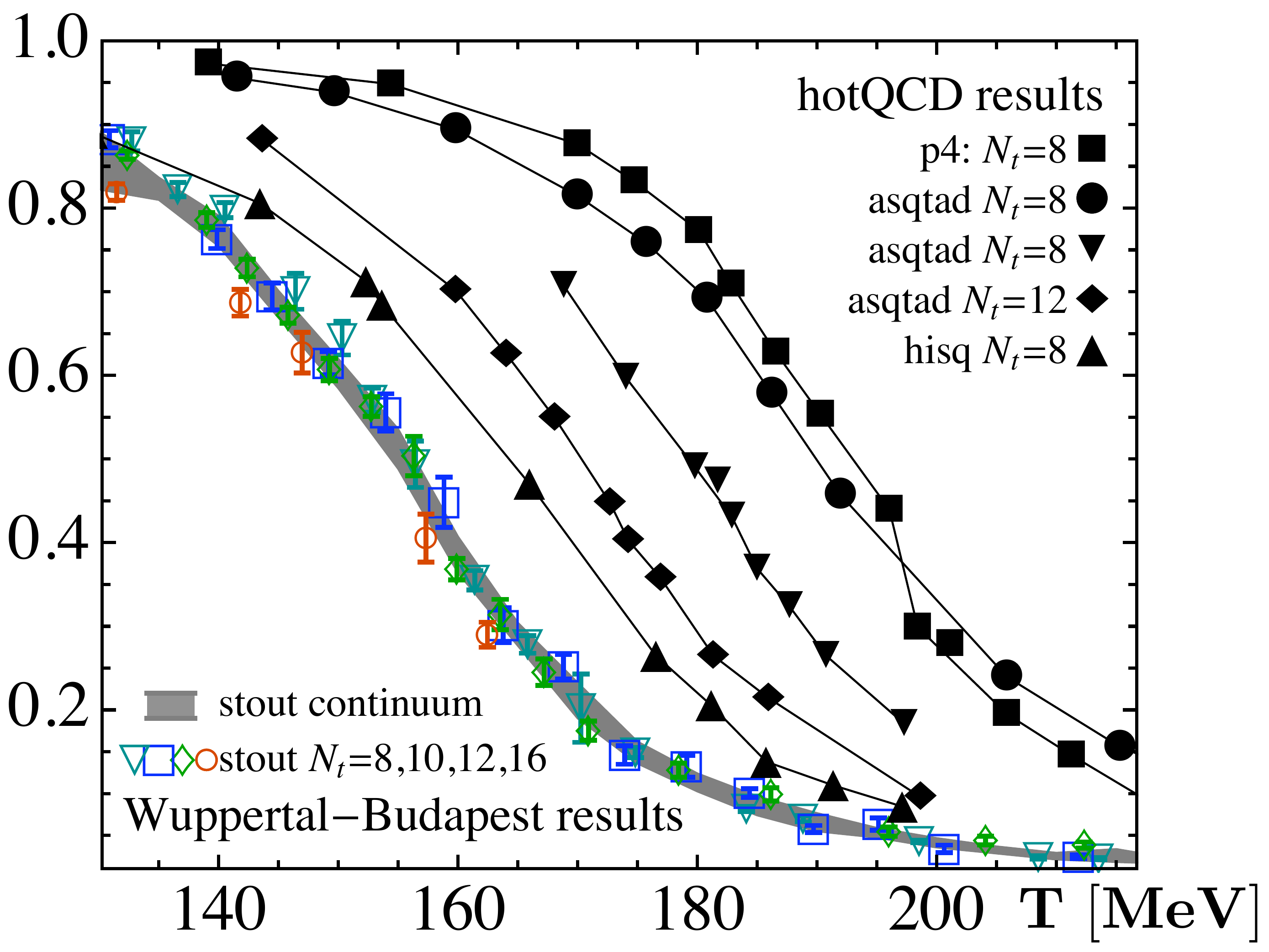}}
\label{fig:5}
\caption{Stepwise improved lattice simulations of the chiral condensate from the HotQCD collaboration\cite{BP2010} in comparison with the Wuppertal-Budapest results\cite{Aoki2006,Bo2010}.}
\end{figure}
%

\subsection{Chiral condensate and Polyakov loop}

The order parameter of spontaneously broken chiral symmetry is the quark condensate, $\langle \bar{\psi} \psi \rangle$.  The  disappearence of the condensate, by its melting above a transition temperature range $T_c$, signals the restoration of chiral symmetry in its Wigner-Weyl realization. The transition from confinement to deconfinement in QCD is likewise controlled by an order parameter, the Polyakov loop. A non-vanishing Polyakov loop $\Phi$ reflects the spontaneously broken $Z(3)$ symmetry characteristic of the deconfinement phase. The Polyakov loop vanishes in the low-temperature, confinement sector of QCD.

One observes that the chiral and deconfinement transitions as shown in Fig.4 are not phase transitions but smooth crossovers, so there is no critical temperature in the strict sense. It is nevertheless useful to define a transition temperature band around the maximum slope of either the condensate $\langle\bar{\psi}\psi\rangle(T)$ or the Polyakov loop $\Phi(T)$. 

Two limiting cases are of interest in this context. In pure gauge QCD, corresponding to infinitely heavy quarks, the deconfinement transition is established as a first order phase transition with a critical temperature of about $270$ MeV. In the limit of massless $u$ and $d$ quarks, on the other hand, the isolated chiral transition appears as a second order phase transition at a significantly lower critical temperature. This statement is based on calculations using Nambu - Jona-Lasinio (NJL) type models which incorporate the correct spontaneous chiral symmetry breaking mechanism but ignore confinement. The step from first or second order phase transitions to crossovers is understood as a consequence of explicit symmetry breaking. The $Z(3)$ symmetry is explicitly broken by the mere presence of quarks with non-infinite masses. Chiral symmetry is explicitly broken by non-zero quark masses. But the challenging question remains whether and how the chiral and deconfinement transitions get dynamically entangled in just such a way that they finally occur within overlapping transition temperature intervals.

\subsection{The non-local PNJL model}

Insights concerning this issue can be gained from a model based on a minimal synthesis of the NJL-type spontaneous chiral symmetry breaking mechanism and confinement implemented through Polyakov loop dynamics. This PNJL model \cite{Fu2003,RTW2006} is specified by the following action:
\begin{eqnarray}
{\cal S} = \int_0^{\beta=1/T}d\tau \int_V d^3x\left[\psi^\dagger\partial_\tau\psi-{\cal H}(\psi,\psi^\dagger,\phi)\right] - {V\over T}\,{\cal U}(\Phi,T)~.
\end{eqnarray}
It introduces the Polyakov loop, 
\begin{equation}
\Phi = N_c^{-1}\,Tr\exp(i\phi/T)~,
\end{equation}
with a homogeneous temporal gauge field, $\phi = \phi_3\lambda_3 + \phi_8\lambda_8\in SU(3)$, coupled to the quarks. The dynamics of $\Phi$ is controlled by a $Z(3)$ symmetric effective potential ${\cal U}$, designed such that it reproduces the equation of state of pure gauge lattice QCD with its first order phase transition at a critical temperature of 270 MeV. The field $\phi$ acts as a potential on the quarks represented by the flavor doublet (for $N_f = 2$) or triplet (for $N_f = 3$) fermion field $\psi$. The Hamiltonian density in the quark sector is 
\begin{equation}
{\cal H} = -i\psi^\dagger(\vec{\alpha}\cdot\vec{\nabla} +\gamma_4\,\hat{m} - \phi)\psi + {\cal V}(\psi,\psi^\dagger)~,
\end{equation}
 with the quark mass matrix $\hat{m}$ and a chiral $SU(N_f)_L\times SU(N_f)_R$ symmetric interaction ${\cal V}$. 

Earlier two-flavor versions of the PNJL model \cite{Fu2003,RTW2006,RRW2007} have still used a local four-point interaction of the classic NJL type, requiring a momentum space cutoff to regularize loops. A more recent version \cite{HRCW2009}, using a non-local interaction, does not need any longer an artificial $NJL$ cutoff. It generates instead a momentum dependent dynamical quark mass, $M(p)$, along with the non-vanishing quark condensate. A further extension to three quark flavors \cite{HRCW2010} includes a $U(1)_A$ breaking term implementing the axial anomaly of QCD. This term is constructed as a non-local generalization of the Kobayashi-Maskawa-'t Hooft $3\times 3$ determinant interaction. 

A basic relation derived in this non-local PNJL model is the gap equation generating the momentum dependent quark mass self-consistently together with the chiral condensate. Its form (written here for the two-flavor case) is reminiscent of the corresponding equation emerging in Dyson-Schwinger approaches to QCD:
\begin{equation}
M(p) = m_0 + 8 N_c G\int{d^4 q\over (2\pi)^4} {\cal C}(p-q){M(q)\over q^2 + M^2(q)}~,
\end{equation}
where $m_0 = m_u = m_d$ is the current quark mass, G is a coupling strength of dimension (length)$^2$ and ${\cal C}(q)$ is a momentum space distribution, the Fourier transform of which represents the range over which the
non-locality of the effective interaction between quarks extends in (Euclidean) 4-dimensional space-time. This distribution is normalized as ${\cal C}(q=0) =1$. Fig. 6 (left) shows ${\cal C}(p)$ in comparison with the ``classic" NJL cutoff by a step function in momentum space. This translates into the dynamical quark mass of Fig. 6 (right). The non-local approach now permits to establish contacts with the high-momentum limit of QCD with its well established behaviour $M(p) \propto -\alpha_s(p)\langle\bar{\psi}\psi\rangle / p^2$ at $p\rightarrow \infty$. At 
$p \lesssim 1$ GeV the distribution ${\cal C}(p)$ is designed to follow lattice QCD and Dyson-Schwinger results, 
or it may be deduced from an instanton liquid model (see Ref.~\citen{HRCW2010} for more details). The non-local PNJL model can indeed be ``derived" from QCD as demonstrated in Refs.~\citen{Ko2010,MP08}.

\begin{figure}[htb]
\begin{minipage}[t]{7cm}
\includegraphics[width=6.5cm]{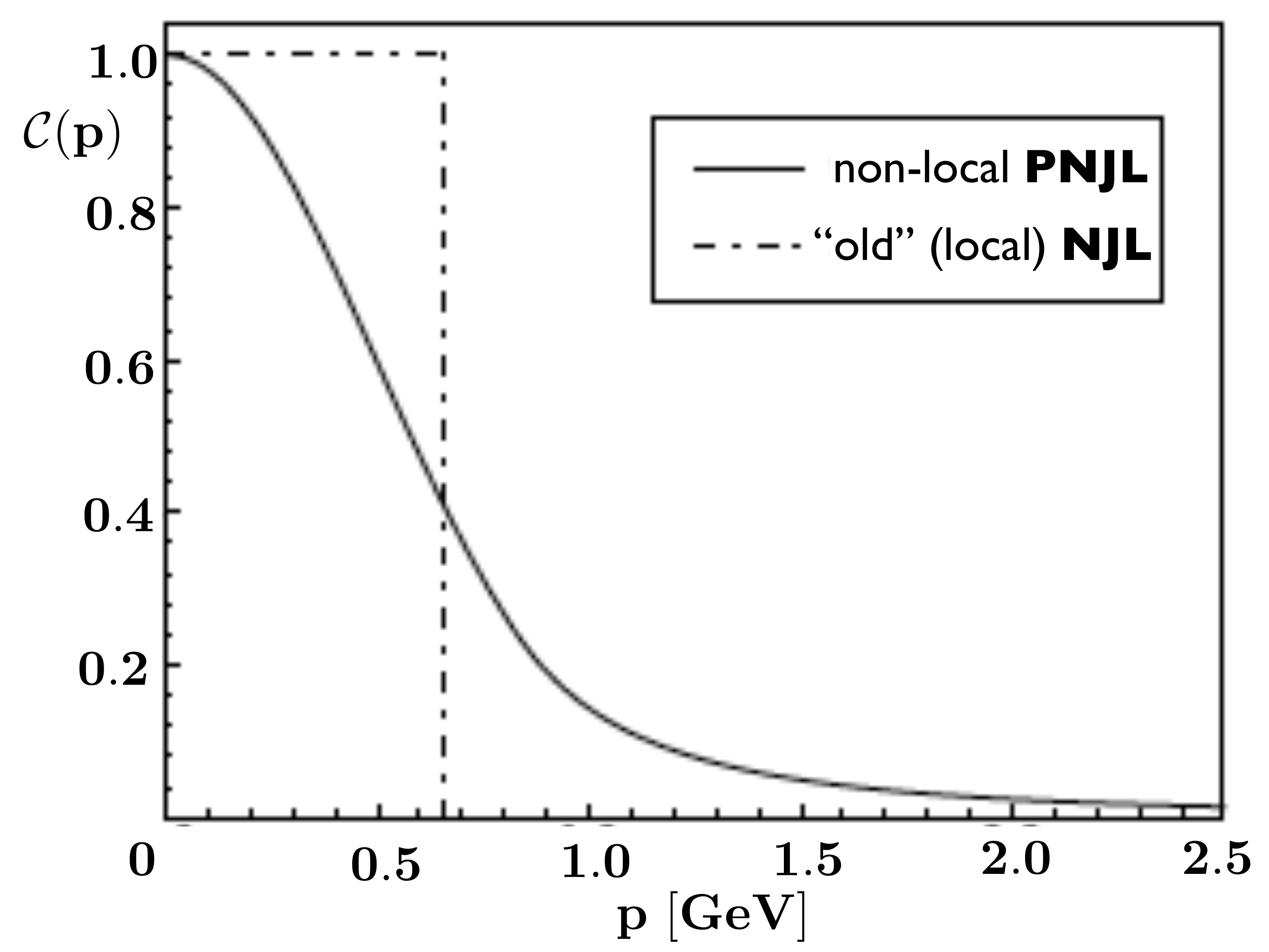}
\end{minipage}
\hspace{\fill}
\begin{minipage}[t]{7cm}
\includegraphics[width=6.5cm]{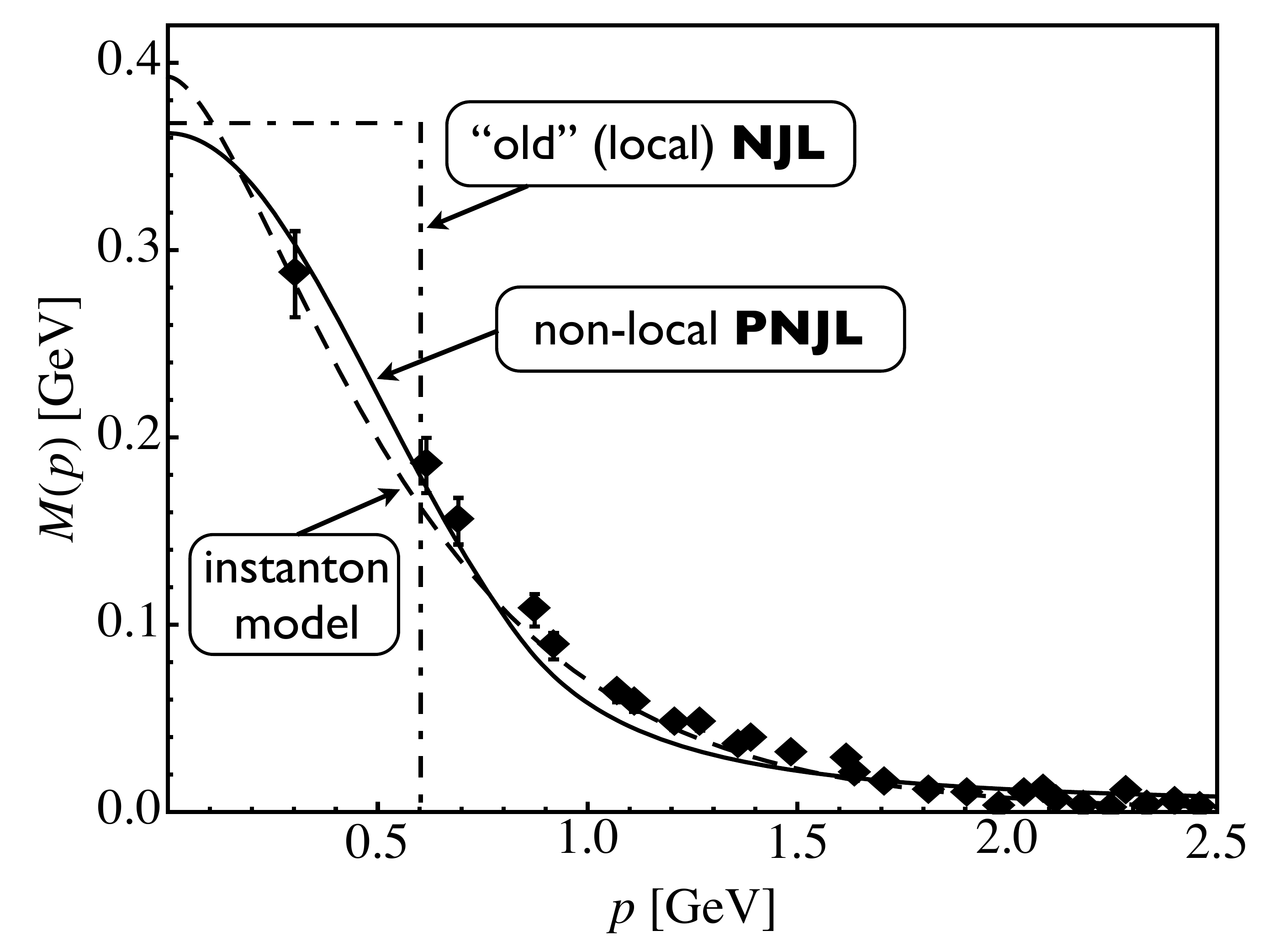}
\end{minipage}
\caption{Left: distribution ${\cal C}(p)$ representing the non-local effective interaction between quarks. The  momentum space cutoff function of the ``old" local NJL model is also shown. Right: dynamical quark mass (solid curve) resulting from the gap equation of the non-local model. Lattice QCD data extrapolated to the continuum and chiral limits \cite{Bo2003} are displayed for comparison, as well as the result deduced from an instanton liquid model. Figures adapted from Ref.~\citen{HRCW2010}\,.} 
\end{figure}
%

\subsection{PNJL thermodynamics}

Once the input is fixed at zero temperature by well-known properties of the pseudoscalar mesons, the thermodynamics of the PNJL model can now be investigated with focus on the symmetry breaking pattern and on the intertwining of chiral dynamics with that of the Polyakov loop. The primary role of the Polyakov loop and its coupling to the quarks is to suppress the thermal distribution functions of color non-singlets, i.e. quarks and diquarks, as the transition temperature $T_c$ is approached from above. Color singlets, on the other hand,  are left to survive below $T_c$. This is seen by analyzing the relevant piece of the thermodynamic potential $\Omega= - (T/V)\ln{\cal Z}$, the one involving the quark quasiparticles:
\begin{eqnarray}
{\Delta\Omega\over T} &\propto& \int {d^3p\over (2\pi)^3}\ln\left[1 + 3\Phi\, e^{-(E_p-\mu)/T} + 3\Phi^*\, e^{-2(E_p-\mu)/T} + e^{-3(E_p-\mu)/T}\right] \nonumber \\
& + & \int {d^3p\over (2\pi)^3}\ln\left[1 + 3\Phi^*\, e^{-(E_p+\mu)/T} + 3\Phi\, e^{-2(E_p+\mu)/T} + e^{-3(E_p+\mu)/T}\right] \,.
\end{eqnarray}
This suppression of color non-singlets in the hadronic phase below $T_c$ should, however, not be interpreted as dynamical confinement. The unsuppressed color singlet three-quark degrees of freedom are not clustered but spread homogeneously over all space. At this stage, nucleons are not treated properly as localized, strongly correlated three-quark compounds plus sea quarks. Thus one cannot expect that such a model describes properly the low-temperature phase of matter at finite baryon chemical potentials around 1 GeV, the domain of nuclear many-body systems.
 
\begin{figure}[t]
\begin{minipage}[t]{7cm}
\includegraphics[width=6cm]{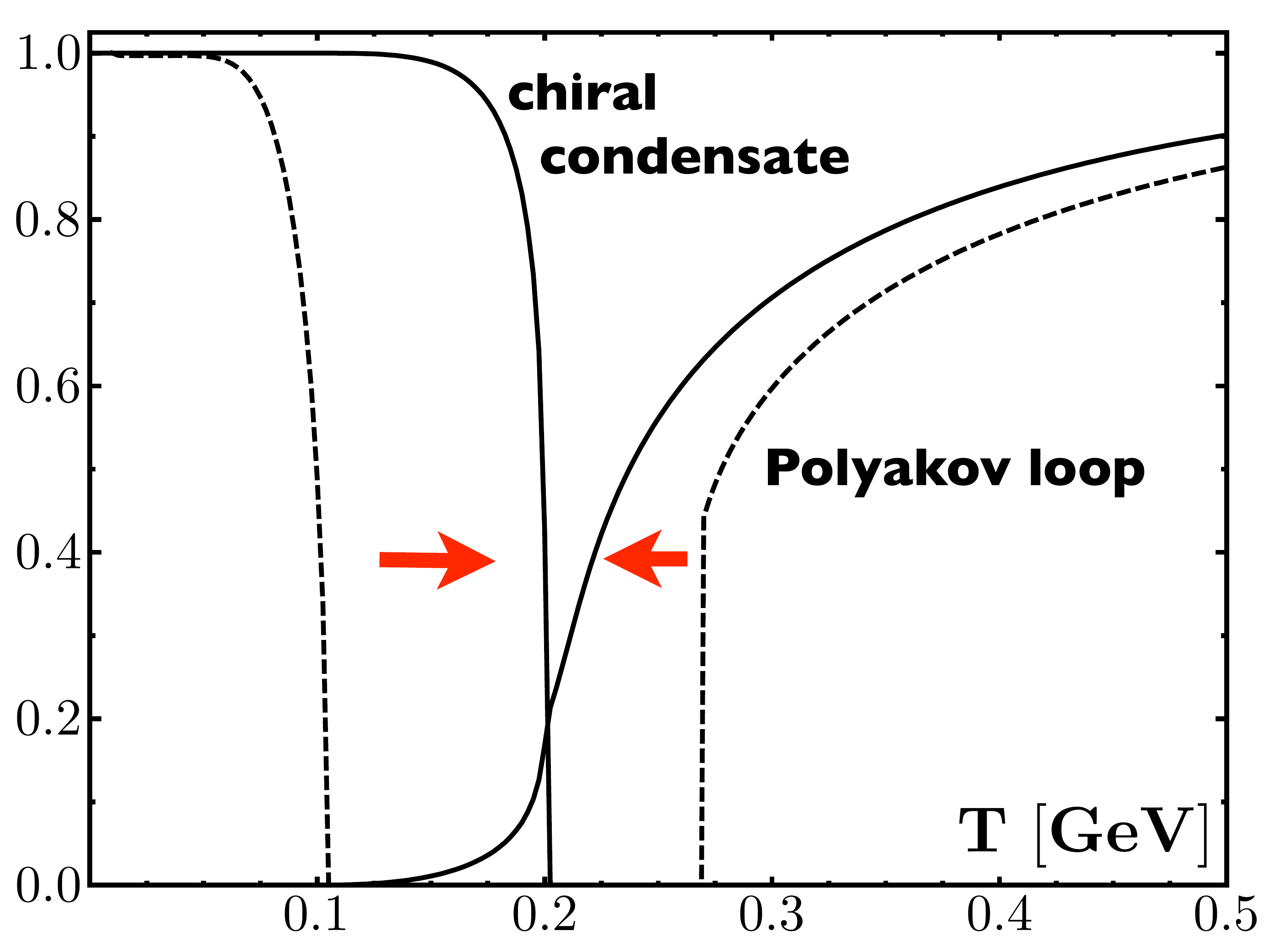}
\end{minipage}
\hspace{\fill}
\begin{minipage}[t]{7cm}
\includegraphics[width=6cm]{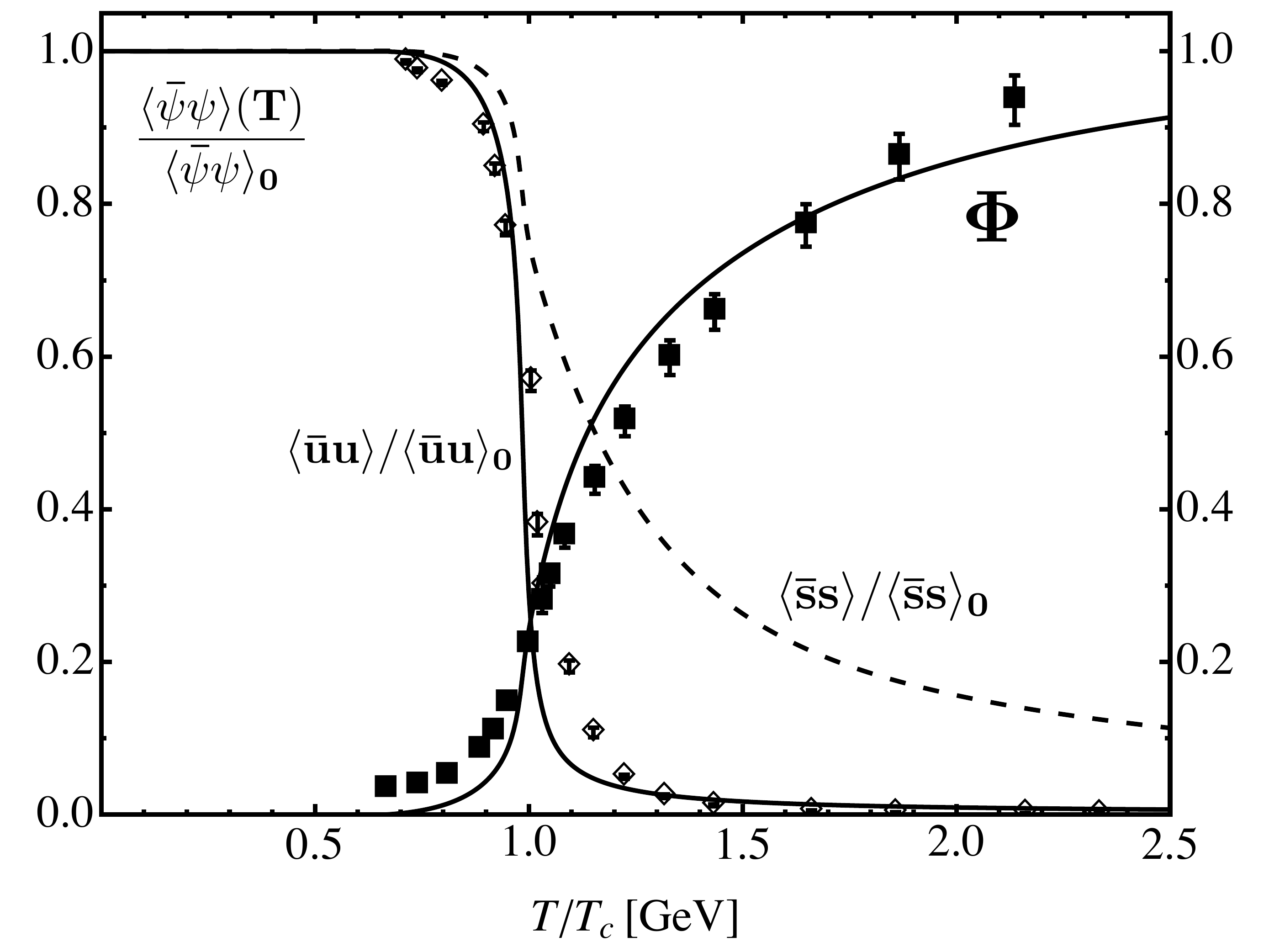}
\end{minipage}
\caption{PNJL model calculations of chiral and deconfinement transitions
\cite{RRW2007,HRCW2009,HRCW2010}. Left: two-flavor PNJL results with dashed curves showing the quark condensate in the chiral limit and the Polyakov loop while they are still decoupled. Coupling the quarks to the Polyakov loop background produces shifts to the corresponding solid curves, i.e. the entanglement of chiral and deconfinement transitions. Right: three-flavor non-local PNJL model results including explicit chiral symmetry breaking by non-zero quark masses $m_{u,d}$ and $m_s$. Also shown are lattice QCD results\cite{Ch2008} for 2+1 quark flavors with almost physical masses.}
\end{figure}
%

At zero baryon chemical potential, a remarkable dynamical entanglement of the chiral and deconfinement transitions is nonetheless verified in the PNJL model,  as demonstrated in Fig.7 (left) for the two-flavor case. In the absence of the Polyakov loop the quark condensate (left dashed line), taken in the chiral limit, shows the expected 2nd order chiral phase transition, but at a temperature way below and far separated from the 1st order deconfinement transition controlled by the pure-gauge Polyakov loop effective potential ${\cal U}$ (right dashed line). Once the coupling of the Polyakov loop field to the quark density is turned on, the two transitions move together and end up at a common transition temperature around 0.2 GeV. The deconfinement transition becomes a crossover (with $Z(3)$ symmetry explicitly broken by the coupling to the quarks), while the chiral phase transition remains 2nd order until non-zero quark masses, $m_{u,d} \simeq 4$ MeV, induce a crossover transition as well.  

It is interesting to compare this PNJL description of the interplay between the chiral and deconfinement transitions, with a scenario based on strong-coupling lattice QCD with inclusion of Polyakov loop dynamics \cite{NMO2010} . Although the starting points and frameworks are quite different in those two approaches, recent studies \cite{NMO2010} arrive at similar results concerning the close correspondence between chiral and deconfinement
transition temperature ranges. 

The right side of Fig.7 shows the three-flavor PNJL result \cite{HRCW2010} together with $N_f=2+1$ lattice data \cite{Ch2008}. A direct comparison is clearly not appropriate but the similarity of the crossover transition patterns is striking, given the simplicity of the model and the fact that these results are derived from a mean-field approach. Further important steps toward systematic investigations beyond mean field are being pursued\cite{CHKW2010}. An important step turns out to be the incorporation of pion degrees of freedom as quark-antiquark collective modes using RPA techniques, consistently expanding beyond the mean field level of the non-local PNJL model\cite{HRCW2009,HRCW2010} . Not surprisingly, thermal excitations of 
Nambu-Goldstone bosons add to the pressure below the transition temperature $T_c$, as seen in Fig. 8 (left), but these modes  disappear rapidly above $T_c$ where the quark quasiparticles and the gluonic degrees of freedom take over. 

Pions also have a pronounced influence\cite{HRCW2010,Hell2010} on the chiral condensate as it approaches the transition region from the hadronic phase. The chiral crossover becomes significantly smoother close to $T_c$ as displayed in Fig. 8 (right). Lattice QCD results\cite{Bo2010} with physical pion masses and realistic extrapolations to the continuum limit show a similar tendency.

\begin{figure}[t]
\begin{minipage}[t]{6cm}
\includegraphics[width=6cm]{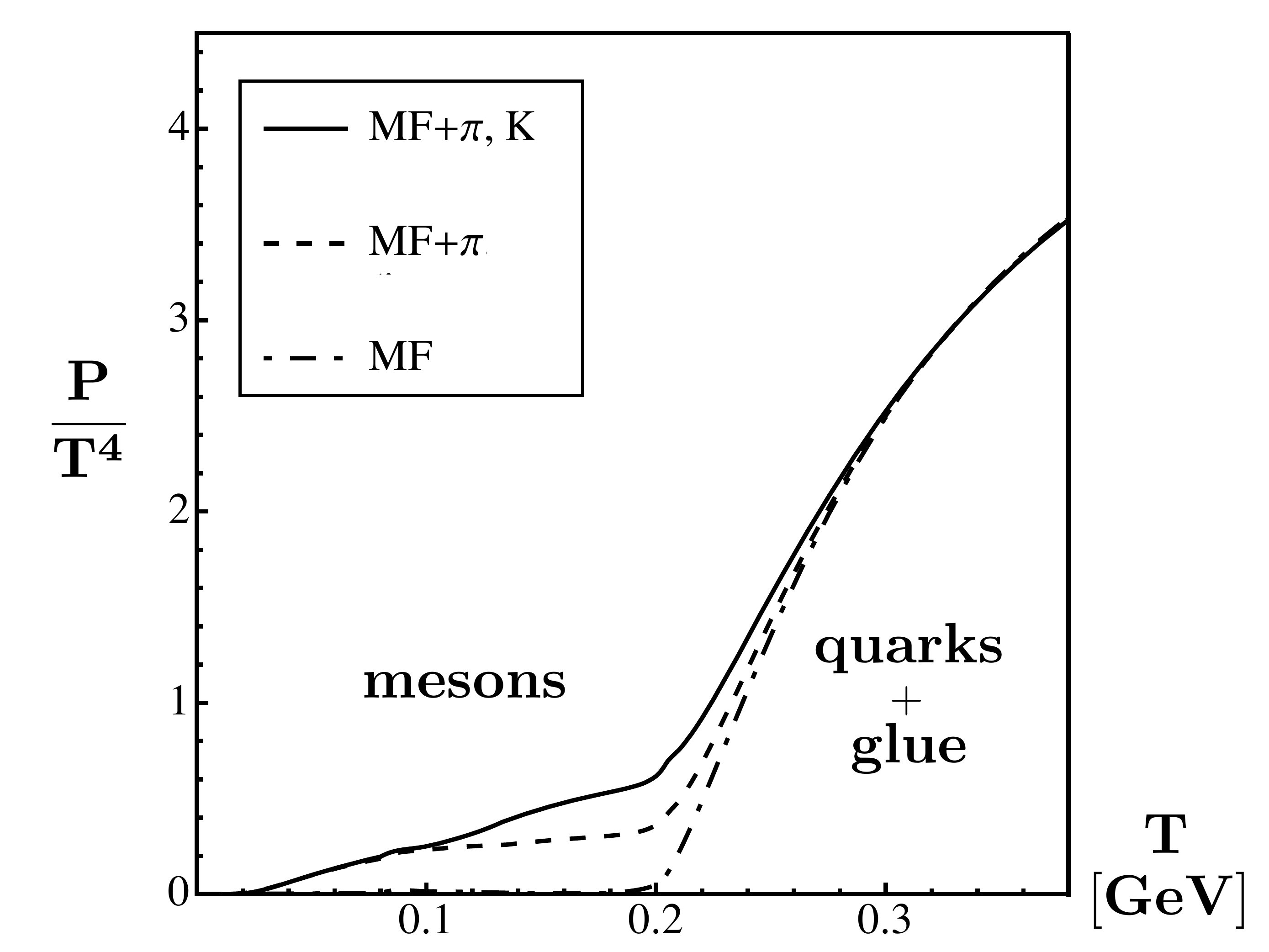}
\end{minipage}
\hspace{\fill}
\begin{minipage}[t]{7cm}
\includegraphics[width=6.5cm]{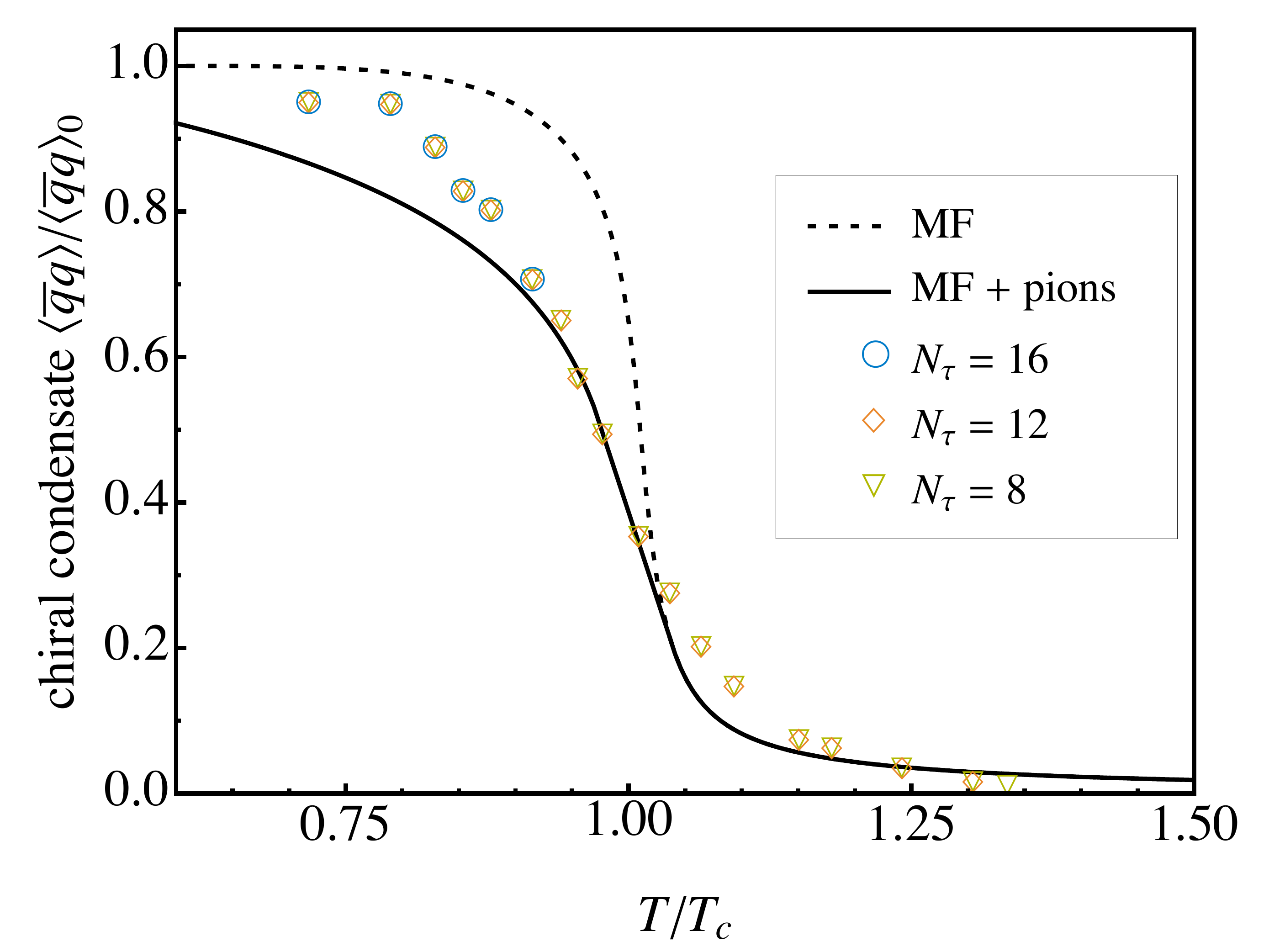}
\end{minipage}
\caption{Effects of mesonic quark-antiquark modes in PNJL model calculations\cite{HRCW2010,Hell2010} of the pressure (left)  and the chiral condensate (right) as functions of temperature. Mean field results (MF) are shown in comparison with those including mesonic contributions (solid curves). Also shown is the chiral condensate from lattice QCD computations of the Wuppertal-Budapest group\cite{Bo2010}.
}
\end{figure}
%

\section{Scenarios at finite baryon density}

\subsection{Modeling the phase diagram of QCD}

Undoubtedly a prime challenge in the physics of strong interactions is the exploration of the QCD phase diagram at non-zero baryon density, extending from normal nuclear matter all the way up to very large quark chemical potentials $\mu$ at which color superconducting phases are expected to occur. PNJL calculations at finite $\mu$ give a pattern of the chiral order parameter in the $(T,\mu)$ plane showing a crossover at small $\mu$ that ends at a critical point. From there on a first-order transition line extends down to a quark chemical potential $\mu \sim$ 0.3 - 0.4 GeV at $T = 0$. 

The typical phase diagram that emerges from the non-local three-flavor PNJL model\cite{HRCW2010} is shown in Fig. 9. The phase to the right of the 1st order transition line, but below the deconfinement boundary that decouples from the chiral transition beyond the critical point, has been named ``quarkyonic"\cite{HLP2008} and is hypothetically assumed to exist until superconducting phases take over at large quark chemical potential. A recent discussion about the possible relationship between such a hypothetical phase and the hadron rates produced in high-energy heavy-ion collisions\cite{An2010} may not yet be conclusive but is pursued with great intensity. 

One must recall at this point that calculations and extrapolations based on PNJL or related approaches still have limited validity at moderate and high baryon densities since they do not properly incorporate baryonic degrees of freedom. They miss the important constraints imposed by the existence of nucleons and nuclear matter. This becomes evident when converting the PNJL phase diagram, Fig. 9, into a plot that translates quark chemical potentials into baryon densities via the derivative of the pressure, $\rho_B = \partial P/\partial\mu_B$, with respect to baryon chemical potential $\mu_B = 3\mu_q$. The 1st order transition line then converts into a coexistence region along the density axis, as illustrated in Fig. 10 (in this case deduced from a local PNJL model calculation\cite{BHW2010}). It becomes immediately evident that this coexistence region encloses what is well known as nuclear
physics terrain with its transition from Fermi liquid to Fermi gas - but now with the ``wrong" quasiparticle degrees of freedom: PNJL quarks rather than interacting nucleons. Consequently, interpretations based on the schematic phase diagram of Fig. 9 along the chemical potential axis should be exercised with caution.

\begin{figure}[htb]
\begin{minipage}[t]{6.5cm}
\includegraphics[width=6.5cm]{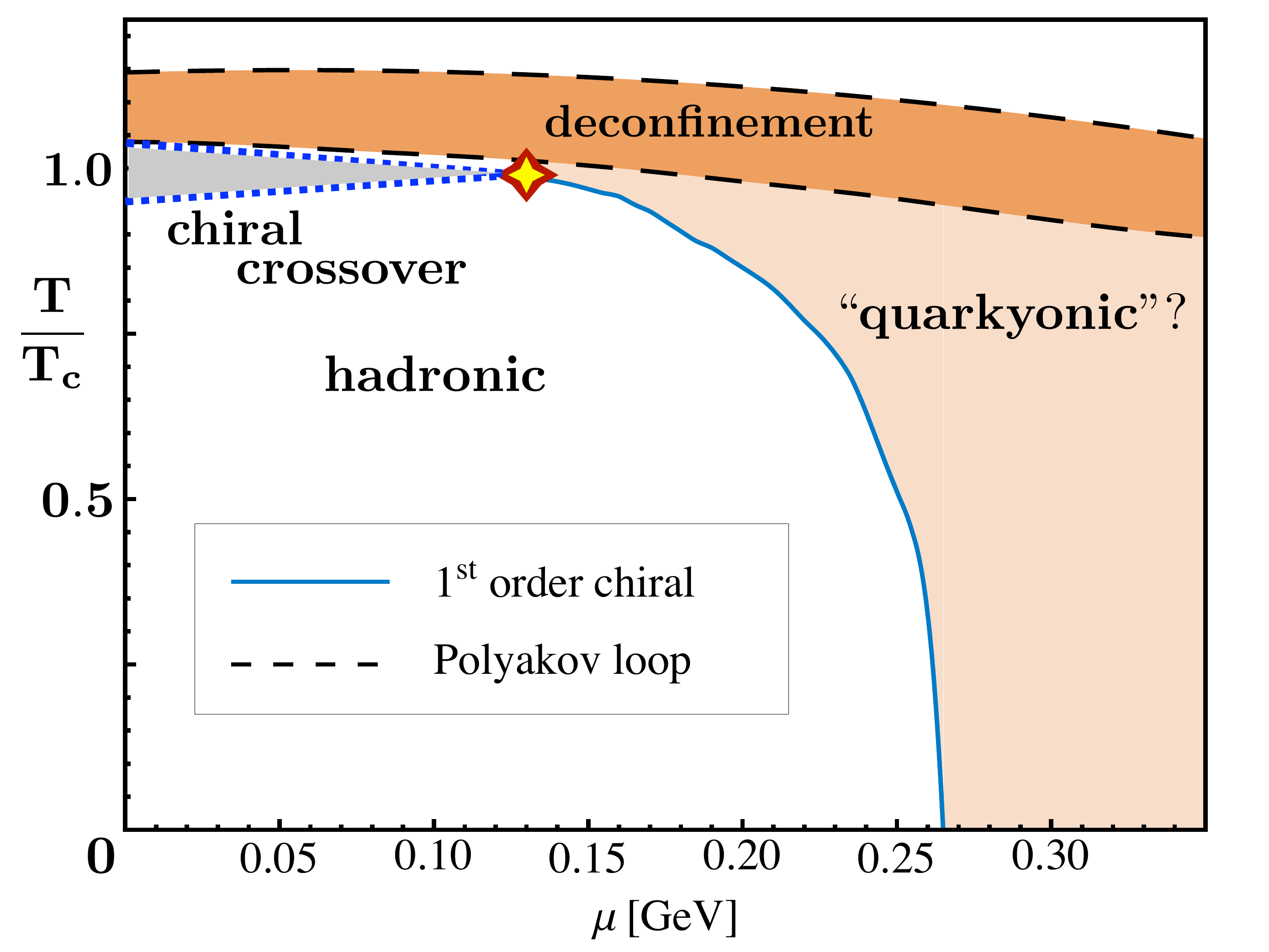}
\caption{Three-flavor non-local PNJL model calculation \cite{BHW2010} of the phase diagram in the plane of temperature and non-strange quark chemical potential. At the critical point the chiral crossover turns into a 1st order transition. The deconfinement transition band represents the crossover range observed in the Polyakov loop.} 
\end{minipage}
\hspace{\fill}
\begin{minipage}[t]{7cm}
\includegraphics[width=7cm]{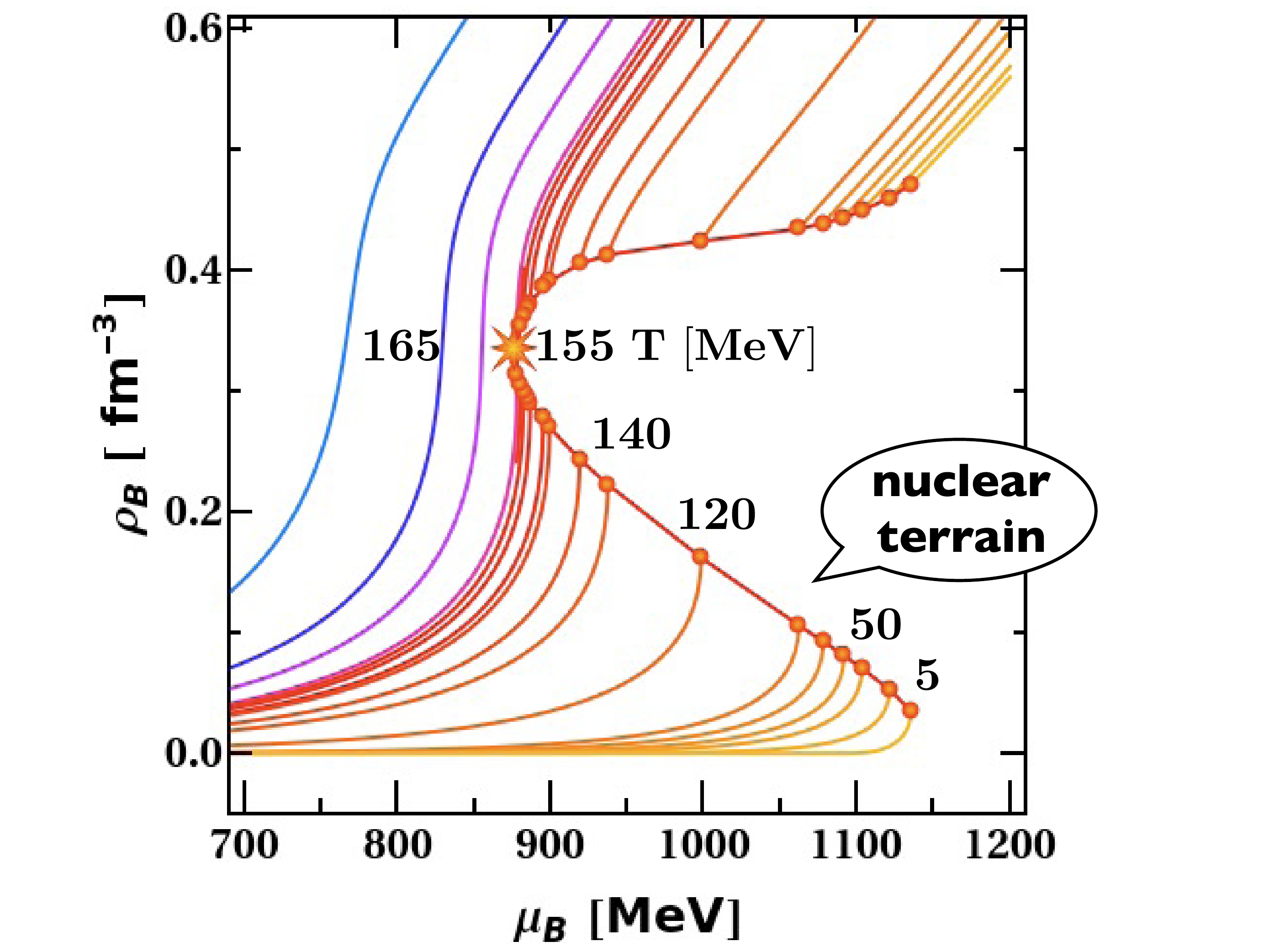}
\caption{Isotherms of the PNJL phase diagram plotted in the plane of baryon density versus baryon chemical potential $\mu_B = 3 \mu_{quark}$. This calculation is based on a local version of the 3-flavor PNJL model\cite{BHW2010}. The domain relevant to nuclear physics is indicated as ``nuclear terrain".} 
\end{minipage}
\end{figure}

\subsection{The quest for the critical point}

Several further important questions are being raised in this context. A principal one concerns the existence and location of the critical point in the phase diagram. Extrapolations from lattice QCD, either by Taylor expansions\cite{Aoki2006} around $\mu =0$ or by analytic continuation from imaginary chemical potential \cite{FP2008}, have so far not reached a consistent conclusion. A second question relates to the sensitivity  of the first order transition line in the phase diagram with respect to the axial $U(1)_A$ anomaly in QCD. This issue has been addressed in Ref.~\citen{YTHB2007} using a general Ginzburg-Landau ansatz for the chiral SU(3) effective action with axial $U(1)$ anomaly. It was pointed out that, depending on details of the axial $U(1)_A$ breaking interaction, a second critical point might appear such that the low-temperature evolution to high density is again just a smooth crossover, or the first order transition might disappear altogether and give way to a smooth crossover throughout. In a three-flavor NJL type realization of this approach\cite{ABHY2010}, it has recently been demonstrated how the interaction between chiral condensate and diquark degrees of freedom, based on the genuine $U(1)_A$ breaking six-point vertex, can open up a smooth transition corridor along the  chemical potential axis in which chiral and diquark condensates may coexist.

\begin{figure}[htb]
\begin{minipage}[t]{6.5cm}
\includegraphics[width=7cm]{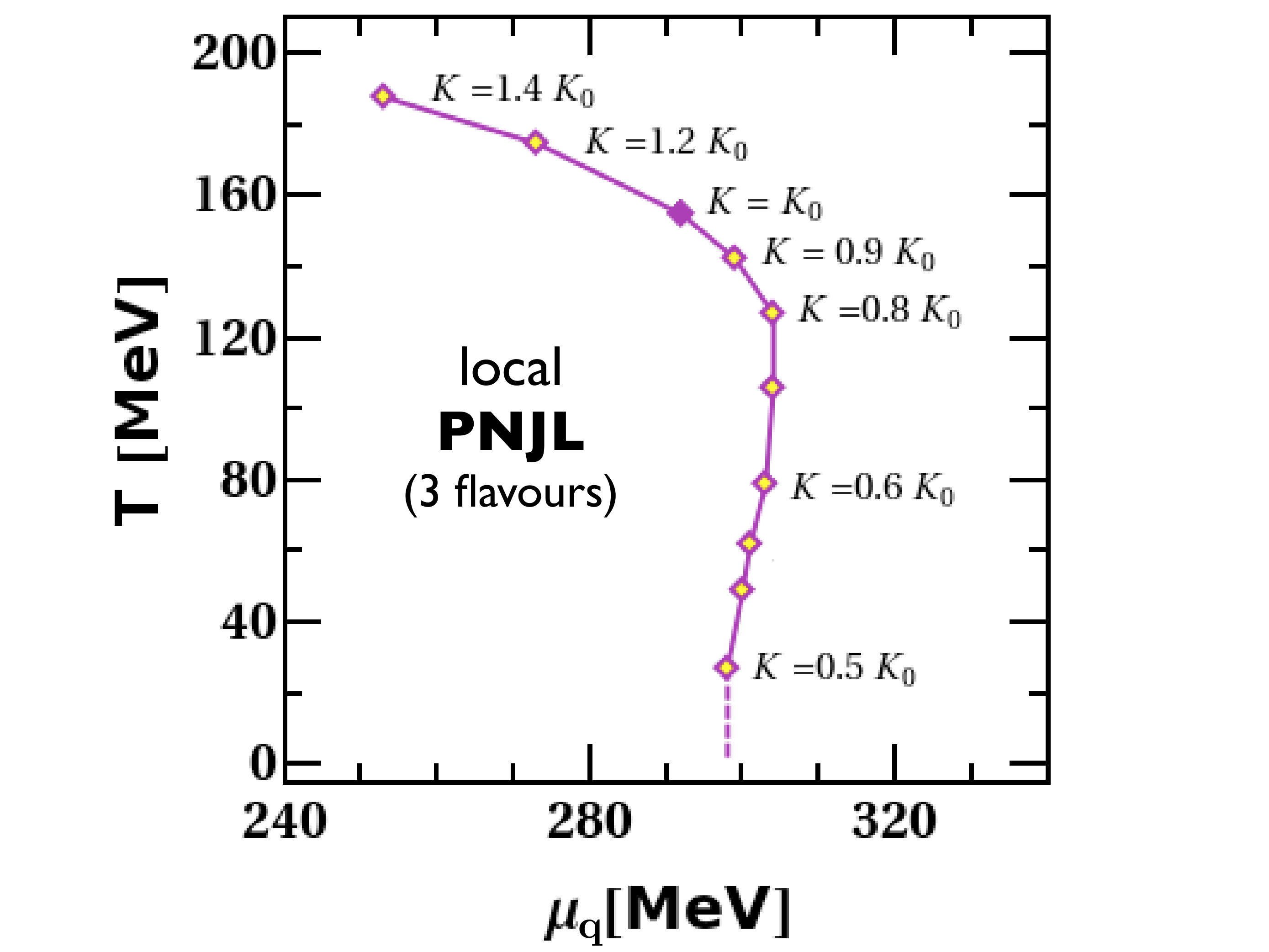}
\caption{Trajectory of the critcal point in the $(T,\mu)$ plane as function of $K$, the strength of the axial $U(1)$ breaking interaction of the 3-flavor PNJL model\cite{BHW2010} . Here $K_0$ is the coupling strength that reproduces the $\eta'$ mass in vacuum.} 
\end{minipage}
\hspace{\fill}
\begin{minipage}[t]{7cm}
\includegraphics[width=7.2cm]{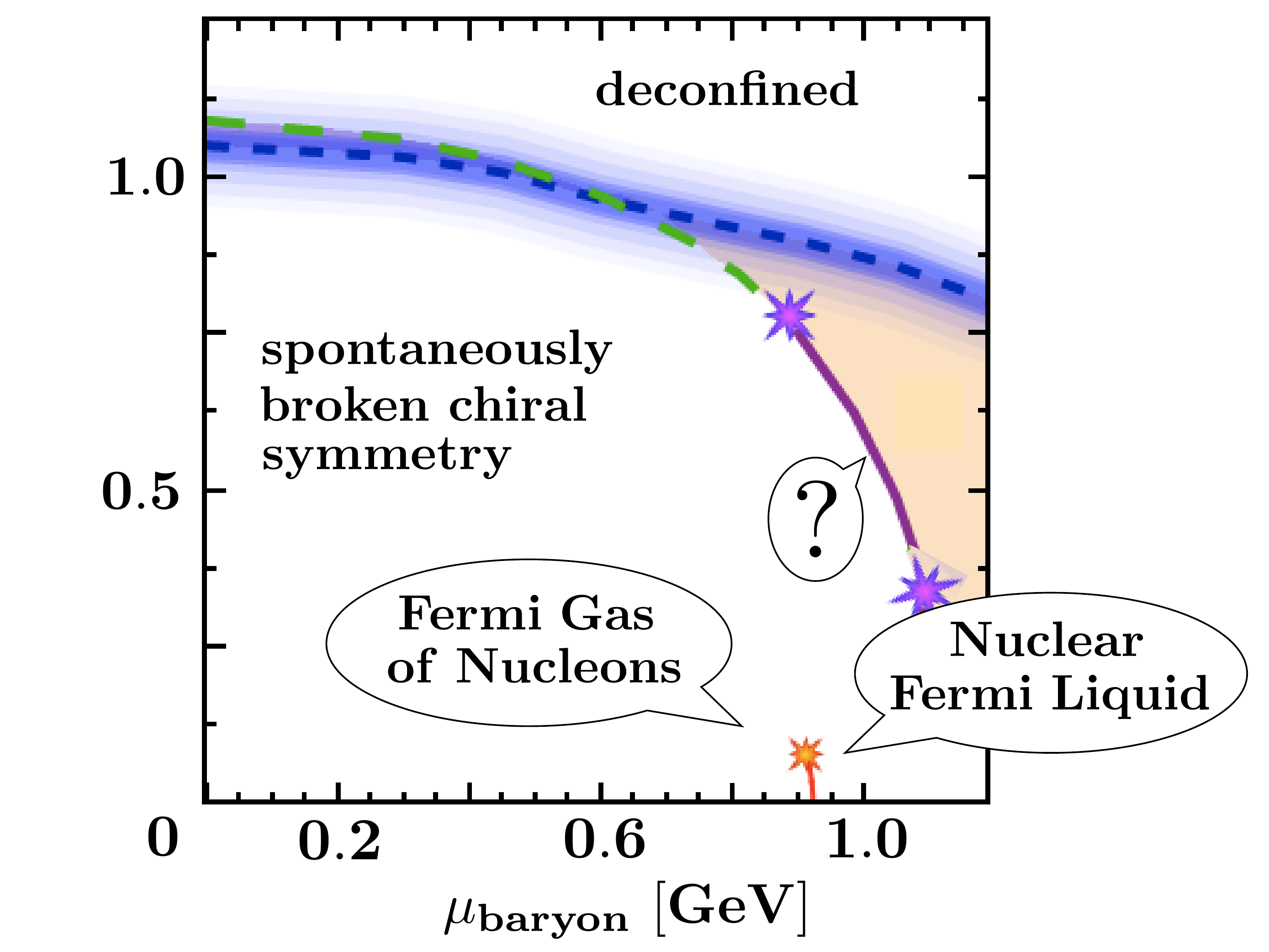}
\caption{A sketch of a PNJL-inspired phase diagram with a smooth crossover corridor along the chemical potential axis.} 
\label{fig:6}
\end{minipage}
\end{figure}

An impression of the explicit dependence of the critical point on the axial anomaly can be obtained using the (local) three-flavor PNJL model with inclusion of a $U(1)_A$ breaking Kobayashi-Maskawa-`t Hooft determinant interaction and varying the coupling strength $K$ of this interaction \cite{Fu2008,BHW2010}. It turns out that the location of the critical point in the phase diagram varies indeed strongly with $K$ and may even disappear altogether below a certain value of $K$ (see Fig. 11). It should be added, though, that this sensitivity to the strength of the axial anomaly appears to be less pronounced in the non-local three-flavor PNJL model\cite{HRCW2010} .

\section{Summary and Outlook}

From the variety of existing model calculations (including those using PNJL models) one might draw the presumably premature conclusion that critical phenomena occur already at a density scale not much higher than that of normal nuclear matter. However, these models are so far not capable of working with the proper degrees of freedom around and below baryon chemical potentials $\sim$ 1 GeV (corresponding to quark chemical potentials around 0.3 GeV). Approaching the corresponding density scale from below, it is obvious that constraints from what we know about the nuclear matter equation of state must be seriously considered (see Fig. 12). This has been the essence of the first part of this presentation. Implementing such nuclear matter constraints (that are well reproduced using the framework of chiral effective field theory) into a consistent picture together with the non-local PNJL model (that has proven to be quite successful at high temperatures) remains a major challenge for the future. 

Finally, in the discussion of high density, low temperature constraints on the QCD equation of state, recent new developments concerning the mass-radius relation deduced from observations of neutron stars in binaries\cite{OBG2010, SLB2010} are potentially of great interest in reducing the set of acceptable equations of state.

\section*{Acknowledgements}
The author gratefully acknowledges the creative atmosphere of the NFQCD programme and the hospitality he experienced at the Yukawa Institute of Theoretical Physics in Kyoto.
Sincere thanks go to my collaborators Nino Bratovic, Marco Cristoforetti, Salvatore Fiorilla, Thomas Hell, Jeremy Holt, Norbert Kaiser and Bertram Klein, whose work has contributed substantially to this report. I am grateful to Kenji Fukushima for many inspiring discussions. This work was partially supported by grants from BMBF, GSI and by the DFG Cluster of Excellence ``Origin and Structure of the Universe".

%

\end{document}